\documentclass[twocolumn,showpacs,prb,preprintnumbers,amsmath,amssymb]{revtex4}
\makeatletter
\def\@dotsep{4,5}
\makeatother

\usepackage{graphicx}
\usepackage{dcolumn}
\usepackage{bm}
\usepackage{subfigure} 

\begin{document}

\preprint{}

\title{Stability of critical bubble in stretched fluid of square-gradient density-functional model with triple-parabolic free energy}

\author{Masao Iwamatsu}
\email{iwamatsu@ph.ns.tcu.ac.jp}
\affiliation{
Department of Physics,
Tokyo City University,
Setagaya-ku, Tokyo 158-8557, Japan 
}%

\author{Yutaka Okabe}
\affiliation{
Department of Physics, 
Tokyo Metropolitan University,
Hachioji, Tokyo 192-0397,
Japan
}

\date{\today}

\begin{abstract}
The square-gradient density-functional model with triple-parabolic free energy, that was used previously to study the homogeneous bubble nucleation [J. Chem. Phys. {\bf 129}, 104508 (2008)], is used to study the stability of the critical bubble nucleated within the bulk under-saturated stretched fluid.  The stability of the bubble is studied by solving the Schr\"odinger equation for the fluctuation.  The negative eigenvalue corresponds to the unstable growing mode of the fluctuation.  Our results show that there is only one negative eigenvalue whose eigenfunction represents the fluctuation that corresponds to the isotropically growing or shrinking nucleus.  In particular, this negative eigenvalue survives up to the spinodal point. Therefore the critical bubble is not fractal or ramified near the spinodal.
\end{abstract}

\maketitle

\section{Introduction}
\label{sec:level1}
The stability of the liquid-vapor interface and its relation to correlations and capillary waves has been studied for more than four decades~\cite{Langer67,Wertheim76,Evans79,Evans81,Hoye87,Bukman97,Varea98}.  The stability of the interface is determined from the eigenvalue problem of the stability matrix~\cite{Wertheim76,Bukman97} or the Schr\"odinger equation~\cite{Langer67,Evans79,Evans81,Bukman97,Varea98} derived from the classical density functional theory~\cite{Evans79}.  It is well recognized that there is always zero eigenvalue which corresponds to the free translation of the planar liquid-vapor interface~\cite{Wertheim76}.  The detailed study of the eigenvalues of the stability matrix and the Schr\"odeinger equation has already been made by Bukman et al~\cite{Bukman97} for the planar interface and by Varea and Robledo~\cite{Varea98} for the curved interface. 

Very recently, a renewed interest in the stability of the liquid-vapor interface has been revived~\cite{Punnathanam03,Uline07,Uline08}.  Uline and Corti~\cite{Uline07,Uline08} have studied the stability of the spherical liquid-vapor interface of the critical bubble~\cite{Punnathanam03,Uline07} and droplet~\cite{Uline08} of the liquid-vapor nucleation, and has shed doubt on the classical view of the so-called classical nucleation theory~\cite{Oxtoby92,Debenedetti96}.  Their work has induced debates~\cite{Lutsko08a,Lutsko08b,Iwamatsu09b,Iwamatsu09c,Lutsko09} on the validity of the concept of the minimum-free energy path on the free-energy landscape of the nucleation.

In this paper, we will use a simple square-gradient density functional theory with a triple-parabolic free energy proposed by Gr\'an\'asy and Oxtoby~\cite{Granasy00}, which has been used to study the various properties of the critical bubble of homogeneous bubble nucleation by the author~\cite{Iwamatsu08b}, to study the stability of the liquid-vapor interface of the critical bubble.  We choose this square-gradient density functional model as it captures the most basic properties of nucleation, yet many physical quantities can be handled analytically. In fact, very recently, Li and Wilemski~\cite{Li03} have compared the results obtained from the accurate density functional theory with the result from the approximate square-gradient theory and found that the two results agree qualitatively well.  In particular, the eigenvalue problem of the stability matrix reduces to the solution of the standard Schr\"odinger equation for a particle in a potential well~\cite{Langer67,Evans81,Bukman97,Varea98} in the square-gradient density functional model which is much simpler than the eigenvalue problem of the stability matrix~\cite{Bukman97,Uline07,Uline08} that consists of roughly $10^{3}$ to $10^4$ elements in the density functional model.  

Although  Varea and Robledo~\cite{Varea98} used a similar square-gradient model with a triple-parabolic free energy to study the stability of curved interface, they were interested in the stability of curved interface in general and paied less attention to the nucleation problem.  In this paper, we will use a more realistic triple-parabolic free energy~\cite{Iwamatsu08b} and pay most attention to the stability of the critical nucleus of homogeneous nucleation including the nucleus near the spinodal point.

In Section II of this paper we will present a short review of the stability of the spherical liquid-vapor interface in the square-gradient density-functional model. In Section III, we will present the numerical results for the stability of the spherical liquid-vapor interface of the critical bubble within the triple-parabolic free energy~\cite{Iwamatsu08b} and discuss the implications of the results in light of the stability of the spherical critical nucleus against non-spherical fluctuations.  Finally Section IV is devoted to the concluding summation.  

\section{Stability of the liquid-vapor interface in the square-gradient density-functional model}
In the square-gradient density-functional model of the fluid~\cite{Evans79,Iwamatsu08b,Yang76,Falls81}, the free energy (grand potential) $W$ of the inhomogeneous fluid, such as the critical bubble in the stretched liquid is written as
\begin{equation}
W=\int \left(\Delta\omega(\phi)+c\left(\nabla \phi\right)^{2}\right) d^{3}{\bf r},
\label{eq:1-1}
\end{equation}
where $c$ is the coefficient of square gradient, and $\Delta \omega\left(\phi\right)$ is the local grand potential density as the function of the local density $\phi\left({\bf r}\right)$.  This form of the free energy is also known as the Cahn-Hilliard model~\cite{Cahn58} or the phase-field model~\cite{Castro03,Iwamatsu08a}.  

The stationary density profile $\phi_{s}$ of the critical nucleus can be determined from the stationary condition of the grand potential written as
\begin{equation}
\delta W\left[\phi\left({\bf r}\right)\right] / \delta \phi\left({\bf r}\right)=0,
\label{eq:1-2}
\end{equation}
which leads to the Euler-Lagrange equation 
\begin{equation}
\frac{\partial \Delta\omega}{\partial \phi}-2c\nabla^{2}\phi\left({\bf r}\right)=0.
\label{eq:1-3}
\end{equation}
Now the stationary profile $\phi_{s}\left({\bf r}\right)$ and the work of formation $W_{s}$ of the critical nucleus can be obtained by solving the differential equation Eq.~(\ref{eq:1-3}). 

The stability of this stationary profile $\phi_{s}$ will be studied from the second variation of the functional $W$, which, after integration by parts, becomes
\begin{equation}
\delta^{2}W=\frac{1}{2}\int \delta\phi\left({\bf r}\right)\left\{\left.\frac{\partial^{2}\Delta\omega}{\partial \phi^{2}}\right|_{\phi_{s}\left({\bf r}\right)}\delta\phi\left({\bf r}\right)-2c\nabla^{2}\delta\phi\left({\bf r}\right)\right\}d^{3}{\bf r},
\label{eq:1-4}
\end{equation}
where $\delta\phi\left({\bf r}\right)$ is a small variation of the order parameter from the stationary profile $\phi_{s}\left({\bf r}\right)$.  Let $\psi_{n}$ be the eigen functions of the Schr\"odinger equation~\cite{Langer67,Evans81,Bukman97,Varea98}
\begin{equation}
-2c\nabla^{2}\psi_{n}\left({\bf r}\right)+v\left({\bf r}\right)\psi_{n}\left({\bf r}\right)=E_{n}\psi_{n}\left({\bf r}\right),
\label{eq:1-5}
\end{equation}  
where
\begin{equation}
v\left({\bf r}\right)=\left.\frac{\partial^{2}\Delta\omega}{\partial \phi^{2}}\right|_{\phi_{s}\left({\bf r}\right)}
\label{eq:1-6}
\end{equation}
plays the role of the potential and $E_{n}$ is the eigenvalue.  Then the order-parameter fluctuation can be expanded as
\begin{equation}
\delta\phi\left({\bf r}\right)=\sum_{n}C_{n}\psi_{n}\left({\bf r}\right),
\label{eq:1-7}
\end{equation}
and the second variation takes the form
\begin{equation}
\delta^{2}W = \frac{1}{2}\sum_{n}E_{n}\left(C_{n}\psi_{n}\left({\bf r}\right)\right)^{2}.
\label{eq:1-8}
\end{equation}

The stability of the stationary solution $\phi_{s}\left({\bf r}\right)$ will be studied from the sign of the eigenvalues $E_{n}$~\cite{Langer67,Varea98}.  In fact, by using the non-conserved order-parameter dynamics~\cite{Castro03,Iwamatsu08a}
\begin{equation}
\frac{\partial \phi}{\partial t}=-\Gamma\frac{\delta W}{\delta \phi\left({\bf r}\right)},
\label{eq:1-9}
\end{equation}
the time evolution of the order parameter $\phi\left({\bf r},t\right)$ near the stationary critical nucleus $\phi_{s}$ is written as
\begin{equation}
\phi\left({\bf r},t\right)=\phi_{s}\left({\bf r}\right)+\delta\phi\left({\bf r},t\right),
\label{eq:1-10}
\end{equation}
and the time evolution of the fluctuation is given by~\cite{Varea98}
\begin{equation}
\delta\phi\left({\bf r},t\right)=\sum_{n}C_{n}\psi_{n}\left({\bf r}\right)\exp\left(-\Gamma E_{n}t\right),
\label{eq:1-11}
\end{equation}
where $\Gamma$ characterizes the time scale of the evolution.  The negative eigenvalue $E_{n}<0$ indicates the unstable growing mode of the order-parameter fluctuation~\cite{Langer67}.  Therefore, the bound state of the Schr\"odigner equation (\ref{eq:1-5}) with negative eigenvalue $E_{n}<0$ corresponds to the unbound growing mode of the fluctuation.  Apparently, the stability analysis is meaningful only near the stationary critical nucleus which satisfies Eq.~(\ref{eq:1-2}).

The stability of critical nucleus can be studied by solving the simple one-body problem of the Schr\"odinger equation (\ref{eq:1-5}) in the simple square-gradient density-functional model.  In contrast, in the original density functional model~\cite{Bukman97,Uline07}, the stability problem leads to the integral equation, which is usually transformed into the eigenvalue problem of the stability matrix with roughly $10^3$ to $10^4$ elements.

\section{Stability of the critical bubble}

\subsection{Triple-Parabolic Model Free Energy}
In order to obtain a more detailed description of the stability of the critical bubble, we chose the triple-parabolic model for the free energy $\Delta\omega\left(\phi\right)$ in Eq.~(\ref{eq:1-1}) used previously to study the scaling properties of the critical bubble~\cite{Iwamatsu08b} originally proposed by Gr\'an\'asy and Oxtoby~\cite{Granasy00}:
\begin{equation}
\Delta \omega(\phi) = \left\{
\begin{array}{ll}
\frac{\lambda_{0}}{2}\left(\phi-\phi_{0}\right)^{2}+\Delta\mu,\;\; & \phi<\phi_{A}, \\
\frac{\lambda_{1}}{2}\left(\phi-\phi_{1}\right)^{2}-\Delta\mu\frac{\phi_{1}-\phi_{0}}{\phi_{2}-\phi_{0}}+\Delta\mu+\omega_{0}, & \\
& \phi_{A}\leq\phi \leq\phi_{B}, \\
\frac{\lambda_{2}}{2}\left(\phi-\phi_{2}\right)^{2},\;\; & \phi_{B}<\phi,
\end{array}
\right.
\label{eq:2-1}
\end{equation}
with $\lambda_{0}, \lambda_{2}>0$ and $\lambda_{1}<0$, which consists of three parabolas centered at the vapor density $\phi_{0}$, and at the free energy barrier $\phi_{1}$, and at the liquid density $\phi_{2}$, which we call "vapor", "spinodal" and "liquid" part of the free energy.  

The parabolic curvatures $\lambda_{0}$ and $\lambda_{2}$ are related to the compressibility of vapor and liquid phases~\cite{Iwamatsu93}, and $\Delta\mu$ is the  free energy difference between the liquid and the vapor. Since we can write $\Delta\mu$ in Eq.~(\ref{eq:2-1}) by $\Delta\mu\phi\simeq \Delta\mu\phi_{2}$, $\Delta\mu$ is, in fact, the reduced chemical potential of the fluid divided by the liquid density $\phi_{2}$. The terminology "over-saturation" is used when $\Delta\mu$ is positive and "under-saturation" when $\Delta\mu$ is negative.  Since a stretched liquid will be considered in this study, we will be mainly concerned with the under-saturation when $\Delta\mu<0$.  From the continuity of the free energy $\Delta \omega(\phi)$, the boundaries $\phi_{A}$ and $\phi_{B}$, as well as the barrier height $\omega_{0}$ and its location $\phi_{1}$ are given as the functions of $\lambda_{0}$, $\lambda_{1}$, $\lambda_{2}$ and the under-saturation $\Delta\mu$~\cite{Iwamatsu08b}.

The liquid spinodal is defined when the metastable liquid phase at $\phi_{2}$ becomes unstable.  This is realized when $\phi_{2}=\phi_{1}$, which leads to the under-saturation for the liquid spinodal~\cite{Iwamatsu08b},
\begin{equation}
\Delta\mu_{\rm spin}=-\frac{1}{2}\frac{\lambda_{0}\lvert\lambda_{1}\rvert}{\lambda_{0}+\left|\lambda_{1}\right|}\left(\phi_{2}-\phi_{0}\right)^{2}
\label{eq:2-2}
\end{equation}
of the stretched liquid. 

In contrast to the previous models~\cite{Shen01,Unger84,Binder84,Wilemski04} where the compressibility diverges continuously as the spinodal is approached, the compressibility remains finite until the spinodal point is reached in our triple-parabolic model as the curvature $\lambda_{0}$ and $\lambda_{2}$ is fixed.

\begin{figure}[htbp]
\begin{center}
\includegraphics[width=0.90\linewidth]{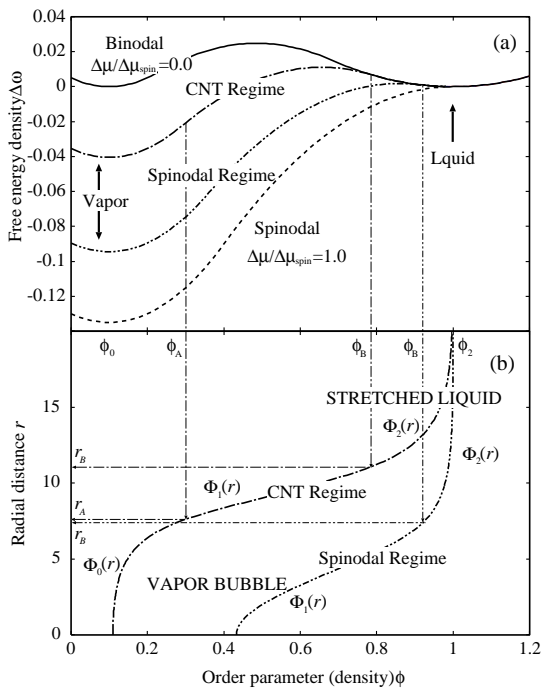}
\end{center}
\caption{
(a) The triple-parabolic free energy from CNT regime near the coexistence to the spinodal regime near the liquid spinodal for the case (i) (Table~\ref{tab:1}). (b) The corresponding critical bubble at the CNT regime ($\Delta\mu/\Delta\mu_{\rm spin}=0.3$) and at the spinodal regime ($\Delta\mu/\Delta\mu_{\rm spin}=0.7$).
}
\label{fig:1}
\end{figure}

In Fig.~\ref{fig:1}(a) we show the typical shapes of the triple-parabolic free energy $\Delta \omega\left(\phi\right)$~\cite{Iwamatsu08b}.   The corresponding density profile $\phi_{s}\left(r\right)$ of the critical bubble is shown in Fig.~\ref{fig:1}(b).  The radii $r_{A}$ and $r_{B}$ are the matching radius that satisfies $\phi\left(r_{A}\right)=\phi_{A}$ and $\phi\left(r_{B}\right)=\phi_{B}$. Since the free energy consists of three parabolas corresponding to the vapor, spinodal and liquid parts, the density profile of the critical bubble consists of three parts that correspond to the three parts of free energy when $\Delta\mu/\Delta\mu_{\rm spin}=0.3$ near the coexistence.  However, as the under-saturation $\Delta\mu$ increases ($\lvert\Delta\mu\rvert$ becomes large) and it approaches the liquid spinodal $\Delta\mu_{\rm spin}$, the matching radius $r_{A}$ vanishes and the density profile consists of only two parts that correspond to the spinodal and the liquid parts (Fig.~\ref{fig:1}(b)).  We use the terminology "CNT regime" for the former regime near the coexistence where the classical nucleation theory (CNT) is expected to be qualitatively correct, and "spinodal regime" for the latter near the spinodal where the spinodal nucleation~\cite{Debenedetti96,Unger84,Binder84,Wilemski04} is expected to occur.


\subsection{Stationary Profile of the Critical Bubble}

\subsubsection{CNT regime}
Density profile of a spherically symmetric critical bubble can be obtained from the Euler-Lagrange equation Eq.~(\ref{eq:1-3}):
\begin{equation}
\frac{1}{r^{2}}\frac{d}{dr}\left( r^{2}\frac{d\phi}{dr}\right)-\frac{1}{2c}\frac{\partial \Delta\omega}{\partial \phi}=0,
\label{eq:3-1}
\end{equation}
which leads to the ordinary differential equations
\begin{equation}
\frac{d^{2}\Phi_{i}}{dr^{2}}+\frac{2}{r}\frac{d\Phi_{i}}{dr}\pm \Gamma_{i}^{2}\Phi_{i}=0,\;\;\;i=0,1,2,
\label{eq:3-2}
\end{equation}
for the three parabolas in Eq.~(\ref{eq:2-1}), where $\Gamma_{i}=\sqrt{\lvert\lambda_{i}\rvert/2c}$ and $\Phi_{i}(r)=\phi(r)-\phi_{i}$, and $+$ sign is used for $i=1$ and $-$ is used for $i=0, 2$ for $\pm$. These differential equations should be solved with appropriate boundary conditions~\cite{Iwamatsu08b}.

The solutions of this Euler-Lagrange equation in Eq.~(\ref{eq:3-2}) for the critical bubble are given by
\begin{eqnarray}
\Phi_{0}(r) &=& \Phi_{0A}r_{A} {\rm csch}\left(\Gamma_{0}r_{A}\right)\sinh\left(\Gamma_{0}r\right)/r,
\nonumber \\
\Phi_{1}(r) &=& 
\csc\left(\Gamma_{1}\left(r_{A}-r_{B}\right)\right) \nonumber \\
&& \times (
-\Phi_{1B}r_{B}\sin\left(\Gamma_{1}\left(r-r_{A}\right)\right) \label{eq:3-3} \\
&&
+\Phi_{1A}r_{A}\sin\left(\Gamma_{1}\left(r-r_{B}\right)\right)
)/r, \nonumber \\
\Phi_{2}(r) &=& \Phi_{2B}r_{B}\exp\left(-\Gamma_{2}r+\Gamma_{2}r_{B}\right)/r, \nonumber
\end{eqnarray}
for the three parts $i=0, 1, 2$ respectively, where
\begin{eqnarray}
\Phi_{0A}=\phi_{A}-\phi_{0}, \nonumber \\
\Phi_{1A}=\phi_{A}-\phi_{1}, \nonumber \\
\Phi_{1B}=\phi_{B}-\phi_{1}, \label{eq:3-4} \\
\Phi_{2B}=\phi_{B}-\phi_{2}. \nonumber
\end{eqnarray}
Finally, the matching radii $r_{A}$ and $r_{B}$ are determined from the simultaneous equations
\begin{eqnarray}
\left.\frac{d\Phi_{2}}{dr}\right|_{r=r_{B}} &=& \left.\frac{d\Phi_{1}}{dr}\right|_{r=r_{B}}, \nonumber \\
\left.\frac{d\Phi_{1}}{dr}\right|_{r=r_{A}} &=& \left.\frac{d\Phi_{0}}{dr}\right|_{r=r_{A}},
\label{eq:3-5} 
\end{eqnarray}
where only $\Phi_{1}$ is a function of both $r_{A}$ and $r_{B}$ (Eq.~(\ref{eq:3-3})).  These simultaneous equations can be solved numerically using standard algorithms such as the Newton-Raphson method.  Even though multiple roots of Eq.~(\ref{eq:3-5}) exist, the smallest $r_{A}$ and $r_{B}$ should be chosen as they corresponds to the nucleus with the lowest free-energy.

\subsubsection{Spinodal regime}

In this case $r_{A}$ becomes zero.   Therefore $i=0$ CNT part (see Fig.~\ref{fig:1}) of the free energy density $\Delta\omega\left(\phi\right)$ in Eq.~(\ref{eq:2-1}) and its solution $\Phi_{0}$ disappears.  Then, the solution for the Euler-Lagrange equation for $\Phi_{2}$ is the same as Eq.~(\ref{eq:3-3}), but the solution for $\Phi_{1}$ now reads
\begin{equation}
\Phi_{1}(r)=\Phi_{1B}r_{B}{\rm csc}\left(\Gamma_{1}r_{B}\right)\sin\left(\Gamma_{1}r\right)/r,
\label{eq:3-6}
\end{equation}
for the critical bubble in the spinodal regime.

In this case, the matching radius $r_{B}$ is simply determined from the equation
\begin{equation}
\left.\frac{d\Phi_{2}}{dr}\right|_{r=r_{B}} = \left.\frac{d\Phi_{1}}{dr}\right|_{r=r_{B}}, 
\label{eq:3-7}
\end{equation}
which is explicitly written as
\begin{equation}
r_{B}\left(\lambda_{2}\Gamma_{1}\cot\left(\Gamma_{1}r_{B}\right)-\lvert\lambda_{1}\rvert\Gamma_{2}\right)=\lambda_{2}+\lvert\lambda_{1}\rvert,
\label{eq:3-8}
\end{equation} 
that does not depend on the under-saturation $\Delta\mu$.  Therefore, the matching radius $r_{B}$ is constant in the spinodal regime.  Again this equation should be solved numerically and the smallest radius $r_{B}$ should be chosen.

In Fig.~\ref{fig:1}(b), we showed the typical density profiles in the CNT regime and in the spinodal regime.  The critical bubble is larger in the CNT regime than in the spinodal regime.  However, the density difference between the inside and the outside of the bubble becomes smaller in the spinodal regime than in the CNT regime.  Correspondingly, the interfacial thickness looks diffuse~\cite{Unger84} as the spinodal is approached.

\begin{figure}[htbp]
\begin{center}
\includegraphics[width=0.80\linewidth]{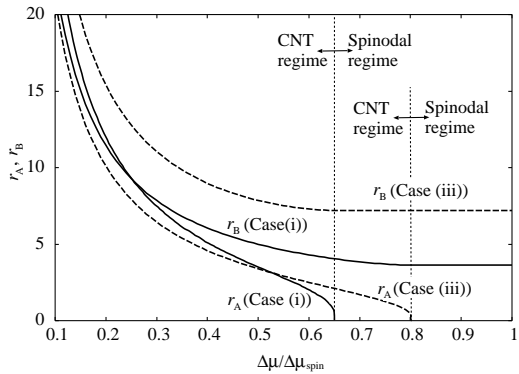}
\end{center}
\caption{
The matching radii $r_{A}$ and $r_{B}$ as functions of the scaled under-saturation $\Delta\mu/\Delta\mu_{\rm spin}$ for the case (i) and (iii) of the free energy parameter listed in Table ~\ref{tab:1}.  In the spinodal regime, $r_{A}$ disappears, and $r_{B}$ becomes constant. 
}
\label{fig:2}
\end{figure}

Figure \ref{fig:2} shows the matching radii $r_{A}$ and $r_{B}$ as functions of the scaled under-saturation $\Delta \mu/\Delta \mu_{\rm spin}$ for the free energy parameter tabulated in Table~\ref{tab:1}. The matching radius $r_{A}$ becomes zero as the under-saturation enters the spinodal regime, while the matching radius $r_{B}$ survives and it becomes constant and independent of the under-saturation $\Delta\mu$ in the spinodal regime as predicted from Eq.~(\ref{eq:3-8}).  

\begin{table}[htbp]
\caption{
Three sets of the free energy parameters in $\Delta\omega\left(\phi\right)$ of Eq.~(\ref{eq:2-1}) used in this work. Case (i) and (ii) were used previously~\cite{Iwamatsu08b} to check the universality of the scaling properties of the critical bubble.}
\label{tab:1}
\begin{center}
\begin{tabular}{c|cccccccc}
\hline
model & $c$ & $\phi_{0}$ & $\phi_{2}$ &  $\lambda_{0}$ & $\lambda_{1}$ & $\lambda_{2}$ & $\lambda_{0}/\left|\lambda_{1}\right|$ & $\lambda_{0}/\left|\lambda_{1}\right|$\\
\hline
case (i)   & 1.0 & 0.1 & 1.0 & 1.0 & -0.5 & 0.3 & 2.0 & 0.6 \\
case (ii)  & 1.0 & 0.1 & 1.0 & 1.0 & -2.0 & 0.9 & 0.5 & 0.45\\
case (iii) & 1.0 & 0.1 & 1.0 & 1.0 & -2.0 & 1.4 & 0.5 & 0.7 \\
\hline
\end{tabular}
\end{center}
\end{table}

\subsection{Eigenvalue problem}
The Schr\"odinger equation Eq.~(\ref{eq:1-5}) for the triple-parabolic model becomes that for a particle in a three-dimensional square well potential:
\begin{equation}
v\left({\bf r}\right) = \left\{
\begin{array}{ll}
\lambda_{0},\;\;\; 0<r<r_{A}, \\
-\left|\lambda_{1}\right|,\;\;\; r_{A}<r<r_{B}, \\
\lambda_{2},\;\;\; r_{B}<r.
\end{array}
\right.
\label{eq:4-1}
\end{equation} 
Figure 3 schematically shows the shape of the potential well.  Depending on the magnitude of the potential barrier $\lambda_{0}$ and $\lambda_{2}$, there are two cases: (a) $\lambda_{0}>\lambda_{2}$ and (b) $\lambda_{2}>\lambda_{0}$.

\begin{figure}[htbp]
\begin{center}
\includegraphics[width=0.80\linewidth]{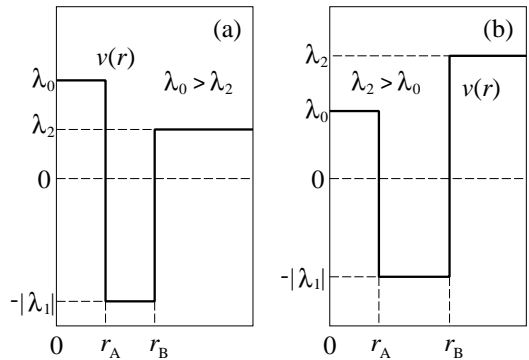}
\end{center}
\caption{
The two cases of the square-well potential $v\left(r\right)$ in Eq.~(\ref{eq:4-1}) as the functions of the radial distance $r$. (a) $\lambda_{0}>\lambda_{2}$, (b) $\lambda_{2}>\lambda_{0}$.  The bound states have eigenvalues $E<\lambda_{2}$.  In case (b) when $\lambda_{2}>\lambda_{0}$, there always exist degenerate bound states $E_{nl}=\lambda_{0}$ for $l\geq 1$. 
}
\label{fig:3}
\end{figure}

Since, we look for the negative eigenvalues, we are interested in the bound state whose eigenvalue $E$ satisfies
\begin{equation}
-\lvert\lambda_{1}\rvert <E\leq \lambda_{2}.
\label{eq:4-2}
\end{equation}
The eigenfunction $\psi\left({\bf r}\right)$ for the bound state of this spherically symmetric Schr\"odinger equation is written as~\cite{Langer67,Schiff68}
\begin{equation}
\psi_{nlm}\left({\bf r}\right)=R_{nl}\left(r\right)Y_{lm}\left(\theta,\phi\right),
\label{eq:4-3}
\end{equation}
using spherical polar coordinates $\left(r,\theta,\phi\right)$, where $Y_{lm}\left(\theta,\phi\right)$ is a spherical harmonic.  The radial part of the eigenfunction $R_{nl}\left(r\right)$ is determined from the radial equation~\cite{Langer67,Schiff68}
\begin{equation}
\frac{1}{r^{2}}\frac{d}{dr}\left(r^{2}\frac{dR}{dr}\right)+\left\{\frac{1}{2c}\left[E-v\left(r\right)\right]-\frac{l(l+1)}{r^2}\right\}R=0.
\label{eq:4-4}
\end{equation}
Differentiating the Euler-Lagrange Eq.~(\ref{eq:3-1}) by $r$, one can easily show that there is always one bound state solution with $E=0$ and $l=1$ whose eigenfunction is given by~\cite{Varea98}
\begin{equation}
R\left(r\right)=d\phi/dr.
\label{eq:4-5}
\end{equation}
Other eigenvalues will be determined numerically.

\subsubsection{CNT regime}

In this regime, $r_{A}$ is finite and we have to solve the Schr\"odinger equation Eq.~(\ref{eq:4-4}) with three wells in Eq.~(\ref{eq:4-1}).  When $\lambda_{2}<\lambda_{0}$  (Fig.~\ref{fig:3}(a)) the radial solution for the bound state with $-\lvert\lambda_{1}\rvert<E\leq \lambda_{2}$ is given by
\begin{equation}
R_{nl}\left({\bf r}\right) = \left\{
\begin{array}{ll}
Aj_{l}\left(i\alpha_{0}r\right),\;\;\; r<r_{A}, \\
Bj_{l}\left(\alpha_{1}r\right)+Cn_{l}\left(\alpha_{1}r\right),\;\;\; r_{A}<r<r_{B}, \\
Dh_{l}^{(1)}\left(i\alpha_{2}r\right),\;\;\; r_{B}<r,
\end{array}
\right.
\label{eq:4-6}
\end{equation}
with $A$, $B$, $C$, and $D$ are constant, and
\begin{eqnarray}
\alpha_{0} &=& \sqrt{\left(\lambda_{0}-E\right)/2c}, \nonumber \\
\alpha_{1} &=& \sqrt{\left(\left|\lambda_{1}\right|+E\right)/2c}, \label{eq:4-7} \\
\alpha_{2} &=& \sqrt{\left(\lambda_{2}-E\right)/2c}, \nonumber 
\end{eqnarray}
where $j_{l}$, $n_{l}$ and $h_{l}^{(1)}$ are the spherical Bessel, Neumann and Hankel functions, respectively~\cite{Schiff68}.

The constants $A$, $B$, $C$ and $D$ are determined from continuity of the wave function Eq.~(\ref{eq:4-6}) and its derivative that leads to the characteristic equation for the energy $E$,
\begin{equation}
\left|
\begin{array}{@{\,}cccc@{\,}}
j_{l}\left(i\xi\right) & -j_{l}\left(\eta\right) & -n_{l}\left(\eta\right) & 0 \\ 
i\xi j_{l}^{'}\left(i\xi\right) & -\eta j_{l}^{'}\left(\eta\right) & -\eta n_{l}^{'}\left(\eta\right) & 0 \\
0 & -j_{i}\left(\zeta\right) & -n_{l}\left(\zeta\right) & h_{l}^{(1)}\left(i\chi\right) \\
0 & -\zeta j_{i}^{'}\left(\zeta\right) & -\zeta n_{l}^{'}\left(\zeta\right) & i\chi h_{l}^{(1)'}\left(i\chi\right) \\
\end{array}
\right|=0
\label{eq:4-8}
\end{equation}
where $j_{l}^{'}\left(x\right)=dj_{l}/dx$ etc. are the derivatives, and
\begin{equation}
\xi = \alpha_{0}r_{A},\;\;\eta=\alpha_{1}r_{A},\;\;\zeta=\alpha_{1}r_{B},\;\;\chi=\alpha_{2}r_{B}.
\label{eq:4-9}
\end{equation}
The roots of Eq. (\ref{eq:4-8}) gives the eigenvalues $E_{n,l} (n=0, 1, 2, \dots)$ for each angular momentum $l (l=0, 1, 2,\dots)$.  Equation (\ref{eq:4-8}) reduces, for example, to 
\begin{eqnarray}
&& \frac{1}{\xi\eta\zeta\chi}\left\{e^{-\chi}\left[\xi\cosh\xi\left(-\zeta\cos\left(\eta-\zeta\right)+\chi\sin\left(\eta-\zeta\right)\right)\right.\right.
\nonumber \\
&&-\left.\left.\eta\sinh\xi\left(\chi\cos\left(\eta-\zeta\right)+\zeta\sin\left(\eta-\zeta\right)\right)\right]\right\}
=0
\label{eq:4-10}
\end{eqnarray}
for $l=0$.  Similar equations can be obtained for $l\geq 1$.  As has been noted in the previous subsection, we always have zero eigenvalues $E_{n=0,l=1}=0$ for $l=1$.

When $\lambda_{2}>\lambda_{0}$ (Fig.~\ref{fig:3}(b)), Eqs.~(\ref{eq:4-6}) to (\ref{eq:4-10}) can be applicable as far as $-\lvert\lambda_{1}\rvert < E\leq\lambda_{0}$.  However, when the eigenvalue $E$ falls within the range $\lambda_{0}\leq E\leq\lambda_{2}$, the radial solution for $r<r_{A}$ in Eq.~(\ref{eq:4-6}) has to be replaced by
\begin{equation}
R_{nl}\left({\bf r}\right) = Aj_{l}\left(\alpha_{0}r\right),\;\;\; r<r_{A},
\label{eq:4-11}
\end{equation}
and $\alpha_{0}$ in Eq.~(\ref{eq:4-7}) is now given by
\begin{equation}
\alpha_{0} = \sqrt{\left(E-\lambda_{0}\right)/2c},
\label{eq:4-12}
\end{equation}
and Eq.~(\ref{eq:4-8}) becomes 
\begin{equation}
\left|
\begin{array}{@{\,}cccc@{\,}}
j_{l}\left(\xi\right) & -j_{l}\left(\eta\right) & -n_{l}\left(\eta\right) & 0 \\ 
\xi j_{l}^{'}\left(\xi\right) & -\eta j_{l}^{'}\left(\eta\right) & -\eta n_{l}^{'}\left(\eta\right) & 0 \\
0 & -j_{i}\left(\zeta\right) & -n_{l}\left(\zeta\right) & h_{l}^{(1)}\left(i\chi\right) \\
0 & -\zeta j_{i}^{'}\left(\zeta\right) & -\zeta n_{l}^{'}\left(\zeta\right) & i\chi h_{l}^{(1)'}\left(i\chi\right) \\
\end{array}
\right|=0,
\label{eq:4-13}
\end{equation}
from which equations similar to Eq.~(\ref{eq:4-10}) are obtained.  The roots of these equations gives the eigenvalues $E_{nl}$ in the range $\lambda_{0}\leq E_{nl}\leq\lambda_{2}$ that are obtained numerically.

It is apparent from Eqs.~(\ref{eq:4-8}) and (\ref{eq:4-13}) that $E=\lambda_{0}$ is always the degenerate eigenvalues for $l\geq 1$ when $\lambda_{2}>\lambda_{0}$ as $\alpha_{0}=0$ and $j_{l}\left(\xi=0\right)=0$ for all $l\geq 1$ when $E=\lambda_{0}$.  Also $E_{n,l=1}=0$ is the root of Eqs.~(\ref{eq:4-8}) and (\ref{eq:4-13}) as $\psi_{n,l=1}=d\phi/dr$ has already satisfied the continuity condition Eq.~(\ref{eq:3-5}) that gives a part of the roots of Eqs.~(\ref{eq:4-8}) and (\ref{eq:4-13}).

\subsubsection{Spinodal regime}
In this case it is a simple textbook problem of a particle confined within a square well potential
\begin{equation}
v\left({\bf r}\right) = \left\{
\begin{array}{ll}
-\left|\lambda_{1}\right|,\;\;\; 0<r<r_{B}, \\
\lambda_{2},\;\;\; r_{B}<r.
\end{array}
\right.
\label{eq:4-14}
\end{equation} 
The radial solution for the bound state is now given by
\begin{equation}
R_{nl}\left({\bf r}\right) = \left\{
\begin{array}{ll}
Aj_{l}\left(\alpha_{1}r\right),\;\;\; 0<r<r_{B}, \\
Bh_{l}^{(1)}\left(i\alpha_{2}r\right),\;\;\; r_{B}<r.
\end{array}
\right.
\label{eq:4-15}
\end{equation}
The constants $A$ and $B$ will be determined from an equation similar to Eqs.~(\ref{eq:4-8}) and (\ref{eq:4-13}) that leads to
\begin{equation}
-i\chi h_{l}^{(1)'}\left(i\chi\right) j_{l}\left(\zeta\right)+\zeta j_{l}^{'}\left(\zeta\right) h_{l}^{(1)}\left(i\chi\right)=0.
\label{eq:4-16}
\end{equation}
This equation can be transformed into
\begin{equation}
\zeta\cot\zeta=-\chi,\;\;\;\zeta^{2}+\chi^{2}=\frac{\left|\lambda_{1}\right|+\lambda_{2}}{2c}r_{B}^{2},
\label{eq:4-17}
\end{equation}
for $l=0$~\cite{Schiff68}, which can be solved graphically.  

Since we are interested in the negative eigenvalues $E<0$, we consider the case when $E=0$ in Eq.~(\ref{eq:4-17}), which leads to
\begin{equation}
\cot\left(\zeta\right)=-\sqrt{\frac{\lambda_{2}}{\left|\lambda_{1}\right|}},
\label{eq:4-18}
\end{equation}
as $\zeta=\sqrt{\lvert\lambda_{1}\rvert/2c}r_{B}$ and $\chi=\sqrt{\lambda_{2}/2c}r_{B}$ when $E=0$.  The roots of Eq.~(\ref{eq:4-18}) $\zeta_{0}$, $\zeta_{1}$, $\dots$, gives the reduced radius $\sqrt{\left|\lambda_{1}\right|/2c}r_{B}$ for which the eigenvalues becomes zero ($E_{n,0}=0$), and are the lower bound for the appearance of one, two, $\cdots$ negative eigenvalues with $l=0$. 

\begin{figure}[htbp]
\begin{center}
\includegraphics[width=0.8\linewidth]{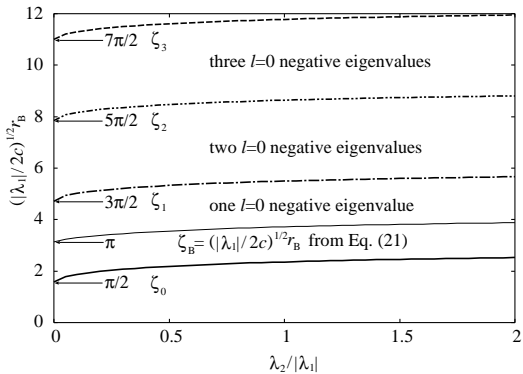}
\end{center}
\caption{
The values $\zeta_{n}=\sqrt{\left|\lambda_{1}\right|/2c}r_{B}, n=0, 1, 2, \dots$ which give the eigenvalues $E_{n,l=0}=0$ with $l=0$.  This eigenvalue $E_{n,l=0}$ becomes negative if the matching radius $r_{B}$ satisfies $\zeta_{n}<\sqrt{\lvert\lambda_{1}\rvert/2c}r_{B}$.  Also shown is the reduced matching radius $\zeta_{B}=\sqrt{\lvert\lambda_{1}\rvert/2c}r_{B}$ determined from the matching condition Eq.~(\ref{eq:3-8}) of the stationary profile of nucleus.
}
\label{fig:4}
\end{figure}

Figure {\ref{fig:4} shows the roots $\zeta_{n}=\sqrt{\lvert\lambda_{1}\rvert/2c}r_{B}$ of Eq.~(\ref{eq:4-18}), which gives the zero eigenvalues $E_{n,l=0}=0$ for $l=0$, as the function of the ratio $\lambda_{2}/\lvert\lambda_{1}\rvert$. When $\lambda_{2}=0$, they are given by $\zeta_{0}=\pi/2$, $\zeta_{1}=3\pi/2$, $\zeta_{2}=5\pi/2$, $\dots$\cite{Schiff68}.  Then, one negative eigenvalue with $l=0$ exist when the matching radius $r_{B}$ satisfies $\zeta_{1}>\sqrt{\lvert\lambda_{1}\rvert/2c}r_{B} \geq \zeta_{0}$, and two eigenvalues exist when $\zeta_{2}>\sqrt{\lvert\lambda_{1}\rvert/2c}r_{B}\geq \zeta_{1}$ etc.  As the barrier $\lambda_{2}$ makes the confinement more effective, the number $\zeta_{n}$ increases as the function of $\lambda_{2}/\lvert\lambda_{1}\rvert$.  Since $\zeta_{0}$ is always smaller than the reduced matching radius $\zeta_{B}=\sqrt{\lvert\lambda_{1}\rvert/2c}r_{B}$ determined from the matching condition Eq.~(\ref{eq:3-8}), we always have one negative eigenvalue and, therefore, the lowest eigenvalues $E_{n=0,l=0}$ is always negative, which corresponds to the isotropically growing/shrinking spherical nucleus~\cite{Langer67,Varea98} with $l=0$.  Other eigenvalues with $l=0$ are always positive as $\zeta_{n}>\zeta_{B}$ for $n=1,2,\dots$.   

Similarly, Eq.~(\ref{eq:4-16}) for $l=1$ can be transformed into
\begin{equation}
\frac{\cot\zeta}{\zeta}-\frac{1}{\zeta^{2}}=\frac{1}{\chi}+\frac{1}{\chi^{2}},\;\;\;\zeta^{2}+\chi^{2}=\frac{\left|\lambda_{1}\right|+\lambda_{2}}{2c}r_{B}^{2},
\label{eq:4-19}
\end{equation}
Again, the condition for the zero eigenvalues $E=0$ with $l=1$ is given by
\begin{equation}
\zeta\cot\zeta-1=\zeta\sqrt{\frac{\left|\lambda_{1}\right|}{\lambda_{2}}}+\frac{\left|\lambda_{1}\right|}{\lambda_{2}}.
\label{eq:4-20}
\end{equation}
It is easy to show that Eq.~(\ref{eq:3-8}) reduces to Eq.~(\ref{eq:4-20}). Therefore the reduced matching radius $\zeta_{B}=\sqrt{\left|\lambda_{1}\right|/2c}r_{B}$ determined from the matching condition Eq.~(\ref{eq:3-8}) of the stationary profile with the lowest free-energy always satisfies Eq.~(\ref{eq:4-20}).  Then, the eigenvalues $E_{n,l}$ for $n=0$ and $l=1$ is always zero ($E_{n=0,l=1}=0$) since $\zeta_{0}=\zeta_{B}$.

\begin{figure}[htbp]
\begin{center}
\includegraphics[width=0.80\linewidth]{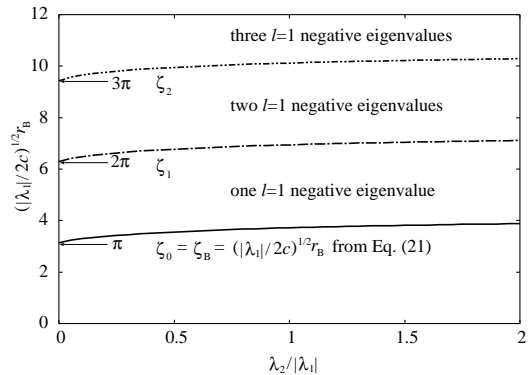}
\end{center}
\caption{
The values $\zeta_{n}=\sqrt{\left|\lambda_{1}\right|/2c}r_{B}, n=0, 1, 2, \dots$ which gives the zero eigenvalue $E_{n,l=1}=0$ for $l=1$, which becomes negative if $\zeta_{n}<\sqrt{\left|\lambda_{1}\right|/2c}r_{B}$.  The reduced matching radius $\zeta_{B}=\sqrt{\left|\lambda_{1}\right|/2c}r_{B}$ determined from the matching condition Eq.~(\ref{eq:3-8}) of the density profile of nucleus coincides with $\zeta_{0}$ ($\zeta_{0}=\zeta_{B}$).
}
\label{fig:5}
\end{figure}

Figure~\ref{fig:5} shows the roots $\zeta_{n}=\sqrt{\lvert\lambda_{1}\rvert/2c}r_{B}$ of Eq.~(\ref{eq:4-20}) which gives the zero eigenvalue $E_{n,l=1}=0$ for $l=1$ as the function of the ratio $\lambda_{2}/\lvert\lambda_{1}\rvert$.  When $\lambda_{2}=0$, they are given by $\pi$, $2\pi$, $3\pi$, $\dots$\cite{Schiff68}.  Again, the roots $\zeta_{n}$ increases as the function of $\lambda_{2}/\lvert\lambda_{1}\rvert$. Since the reduced matching radius $\zeta_{B}=\sqrt{\lvert\lambda_{1}\rvert/2c}r_{B}$ determined from the matching condition Eq.~(\ref{eq:3-8}) coincides with $\zeta_{0}$, the lowest energy $E_{n=0,l=1}$ with $n=0$ and $l=1$ is always zero ($E_{n=0,l=1}=0$).  

If the matching radius $r_{B}$ ($\zeta_{B}$) is erroneously chosen from the multiple roots of Eq.~(\ref{eq:3-8}) such that the excited state with $n=1$ and $l=1$ has zero eigenvalue $E_{n=1,l=1}=0$ ($\zeta_{1}=\zeta_{B}$), then not only the $l=0$ ground-state eigenvalues $E_{n=0,l=0}$ but also the excited-state eigenvalues $E_{n=1,l=0}$ and $E_{n=0,l=1}$ become negative from Figs.~\ref{fig:4} and \ref{fig:5}.  Therefore any numerical error in the determination of the stationary profile $\phi_{s}\left({\bf r}\right)$ of the critical nucleus could result in the multiple negative eigenvalues appearing.

\subsection{Numerical example and discussions}

In order to study the stability problem of the critical bubble, we have solved Eqs.~(\ref{eq:4-8}), (\ref{eq:4-13}) and (\ref{eq:4-16}) to calculate the full spectrum of the bound-state eigenvalues with $l\leq 2$ as the function of the scaled under-saturation $\Delta\mu/\Delta\mu_{\rm spin}$.  Figure \ref{fig:6} shows the bound-state eigenvalues $E$ for the case (i) of Tab.~\ref{tab:1}.  Since $\lambda_{0}>\lambda_{2}$ (Tab.~\ref{tab:1}), the square-well potential has the shape shown in Fig.~\ref{fig:3}(a). The continuum states have the energy $E/\left|\lambda_{1}\right| > \lambda_{2}/\left|\lambda_{1}\right|=0.6$.  These continuum states describe the capillary-like waves induced around the spherical surface of nucleus.  Since this continuum state starts from $E=\lambda_{2}$, there always exist minimum energy $\lambda_{2}$ to excite capillary wave.  

\begin{figure}[htbp]
\begin{center}
\includegraphics[width=0.85\linewidth]{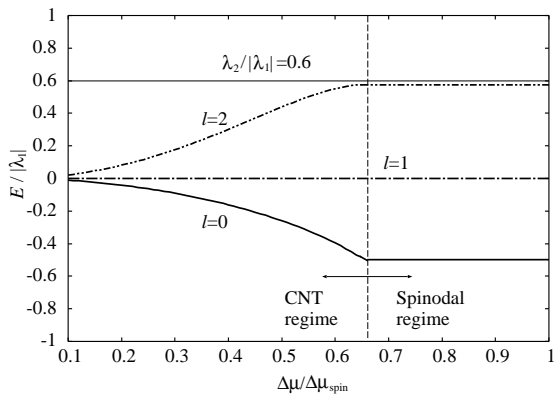}
\end{center}
\caption{
Three bound-state eigenvalues with $l=0,1,2$ as functions of the scaled under-saturation $\Delta\mu/\Delta\mu_{\rm spin}$ for the case (i) of Tab.~\ref{tab:1}.  Only the ground state with $l=0$ has a negative eigenvalues.  The lowest eigenvalues with $l=1$ is always zero.  There is only one negative eigenvalues with $l=0$ from the CNT regime to spinodal regime up to the spinodal point.
}
\label{fig:6}
\end{figure}

There is only one bound state for each $l=0, 1, 2$ in Fig.~\ref{fig:6}.  These eigenvalues depend on the under-saturation $\Delta\mu$ in the CNT regime as the radius $r_{A}$ and $r_{B}$ of the potential well depend on the under-saturation (Fig.~\ref{fig:2}).  Only the ground state eigenvalue for $l=0$ becomes negative and decreases while the ground state eigenvalues for $l\ge 2$ increase as the absolute magnitude $\lvert\Delta\mu\rvert$ is increased toward the spinodal. A similar behavior of the eigenvalues with different $l$ was observed in the numerical results of Varea and Robledo~\cite{Varea98}.  However, the eigenvalues become constant in the spinodal regime in our model as the matching radius $r_{B}$ becomes constant in this regime. Only the ground state with $l=0$ has the negative eigenvalues that describes the growing/shrinking nucleus~\cite{Langer67,Varea98} that preserves the spherical symmetry.  This single negative eigenvalue survives up to the spinodal point.  The lowest eigenvalues with $l=1$ is always zero which means that the deformation with the form 
\begin{equation}
\delta\phi\left({\bf r}\right) \propto \left\{
\begin{array}{l}
\cos\theta\left(d\phi_{s}/dr\right) \\
\sin\theta\left(d\phi_{s}/dr\right) \\
\end{array}
\right.
\label{eq:5-1}
\end{equation}
does not cost energy~\cite{Varea98} as $Y_{l=1,m}(\theta,\phi)\propto \cos\theta, \sin\theta$~\cite{Schiff68}.  This deformation, in fact, describes the translation of the center of spherical nucleus~\cite{Caroli92}, and, therefore, deos not cost energy.

It is possible to choose larger $r_{B}$ ($\zeta_{B}$) from Eq.~(\ref{eq:3-8}) that corresponds, for example, to $\zeta_{1}$ in Fig.~\ref{fig:5}.  However, this false stationary profile has an artificial density oscillation that can be anticipated from Eq.~(\ref{eq:3-6}), and has a higher free-energy.  Figures \ref{fig:4} and \ref{fig:5} also indicate that this false stationary state is unstable against non-spherical growing mode as there will be an extra negative eigenvalue $E_{n=0,l=1}$ with $l=1$.

\begin{figure}[htbp]
\begin{center}
\includegraphics[width=0.85\linewidth]{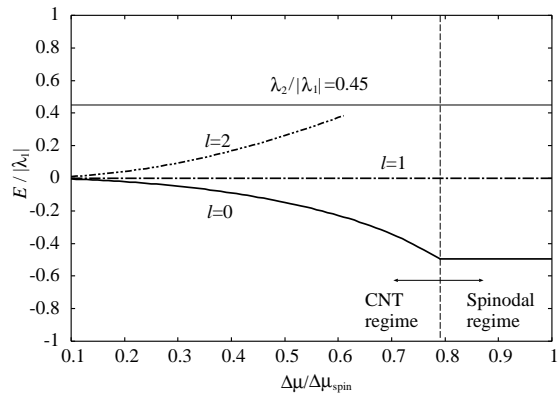}
\end{center}
\caption{
The same as Fig.~\ref{fig:6} for the case (ii) of Tab.~\ref{tab:1}.  Now the bound state for $l=2$ disappears before the spinodal regime is reached as the barrier $\lambda_{2}$ is low.
}
\label{fig:7}
\end{figure}

Figure~\ref{fig:7} shows the bound state energy for the case (ii) of Tab.~\ref{tab:1}.  This is also the case shown in Fig.~\ref{fig:3}(a).  Now the continuum state starts from $E/\lvert\lambda_{1}\rvert>0.45$.  The $l=2$ bound state increases its energy as the under-saturation $\Delta\mu$ is increased and finally it disappears into the continuum state as the confinement $\lambda_{2}$ is weak.  Again there is only one negative eigenvalue with $l=0$ and one zero eigenvalue with $l=1$ which can survive up to the spinodal point.

Figure~\ref{fig:8} shows the bound state energy for the case (iii) of Table~\ref{tab:1}.  In this case, the square-well potential has the shape shown in Fig.~\ref{fig:3}(b) as $\lambda_{2}>\lambda_{0}$ (Tab.~\ref{tab:1}).  In this case we can expect more complex energy diagram for the bound state.  In particular, the degenerate bound states with $E=\lambda_{0}$ appear for all $l\geq 1$ at $E/\left|\lambda_{1}\right|=0.5$  in the CNT regime.  These bound states cannot survive in the spinodal regime as the inner barrier with $E=\lambda_{0}$ disappears in this regime.  Once again there is only one negative eigenvalue with $l=0$ and one zero eigenvalue with $l=1$ up to the spinodal point. 

\begin{figure}[htbp]
\begin{center}
\includegraphics[width=0.85\linewidth]{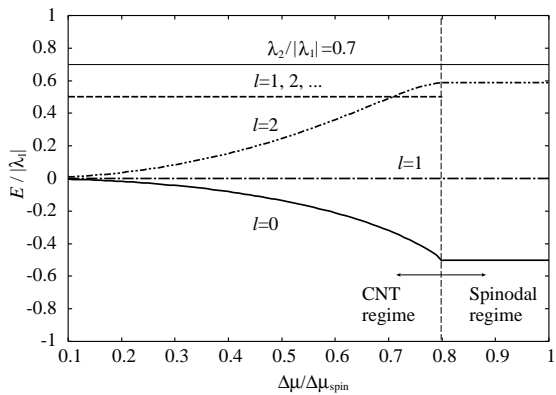}
\end{center}
\caption{
The same as Fig.~\ref{fig:6} for the case (iii) of Tab.~\ref{tab:1}.  In this case new bound-states with the energy $E/\left|\lambda_{1}\right|=0.5$ appears for all $l\ge 1$ in the CNT regime.  
}
\label{fig:8}
\end{figure}

Therefore, the stationary profile of the critical bubble, which satisfies the matching condition of the density profile Eqs.~(\ref{eq:3-5}) and (\ref{eq:3-7}) and has the lowest free-energy, has only one unstable growing mode of the fluctuation with spherical symmetry.  The time scale $\tau$ of the growth or shrinkage is given by the inverse of the absolute magnitude $\lvert E_{n=0,l=0}\rvert$ of the negative eigenvalue with $n=0$ and $l=0$ from Eq.~(\ref{eq:1-11}), that is roughly given by $\tau\sim 1/\left(\Gamma E_{n=0,l=0}\right)\sim 2/\left(\Gamma\lvert\lambda_{1}\rvert\right)$ in the spinodal regime (see Fig.~\ref{fig:6} to \ref{fig:8}).  Therefore the curvature $\lvert\lambda_{1}\rvert$ at the top of the barrier of the free-energy $\Delta\omega\left(\phi\right)$ in Fig.~\ref{fig:1} plays a crucial role in determining the timescale of how fast the saddle point is crossed once the critical bubble is formed.  Of course, the time scale of nucleation that is the time necessary to form a critical bubble is given by the nculeation rate, which is determined by the work of formation of critical bubble~\cite{Punnathanam03,Uline07,Uline08,Lutsko08a,Lutsko08b,Iwamatsu08b}. 

This conclusion persists up to the spinodal point. A diverging compressibility $\lambda_{2}\rightarrow 0$ as $\Delta\mu \rightarrow \Delta\mu_{\rm spin}$ will not affect the timescale of nucleation even near the spinodal.  Also, any fluctuation other than the spherically growing mode that correspond to the negative eigenvalues $E_{n=0,l=0}$ with $n=0$ and $l=0$ stays stable up to the spinodal point. Therefore any structural anomaly~\cite{Klein90,Monette92,Shen01} near the spinodal will start during the growing stage after the nucleation rather than at the nucleation stage of nucleus.  Also this anomaly would be the effect of the interaction of multiple nuclei, such as coalescence or coarsening~\cite{Lifshitz61,Monette92,Iwamatsu99,Yamamoto10} during the growing stage rather than the effect of the instability at the nucleation stage. Since we look at the stationary critical nucleus, the coupling of the order parameter to the diffusion and the heat flow is beyond the scope of the present work.  These effects will also play crucial role in the structural anomaly of nucleus during the growing stage after the nucleation~\cite{Mullins63,Caroli92,Wang09}

\section{Conclusion}
In this study, a square-gradient density-functional model with a triple-parabolic free energy was used to study the stability of the critical bubble of homogeneous bubble nucleation.  By using this square-gradient model~\cite{Varea98,Iwamatsu08b} instead of the original density functional model~\cite{Punnathanam03,Uline07,Uline08}, the stability problem was reduced from the eigenvalue problem of the matrix with roughly $10^{3}$ to $10^{4}$ elements to the text-book problem of the eigenvalues of the Schr\"odinger equation that describes a particle confined within a square-well potential.  The negative eigenvalues of the bound state of the Schr\"odinger equation represents the growing mode of the fluctuation and its magnitude determines the time scale of this growing fluctuation.  We found that there is only one negative eigenvalue that corresponds to the spherically growing/shrinking mode of nucleation.  Our result confirmed the conclusion theoretically predicted~\cite{Langer67} and numerically obtained~\cite{Varea98} by other authors that the critical nucleus is located at the saddle point of the free-energy landscape~\cite{Iwamatsu09c}.    

Furthermore, we have confirmed that this single negative eigenvalue persists up to the spinodal point. Therefore no fractal or ramified structure~\cite{Klein90} is expected at the nucleation stage.  Also, it is clear from our analysis that the stability analysis using the Schr\"odinger equation or the stability matrix is limited only for the critical nucleus at the saddle point in the free-energy landscape.  In order to study the embryonic bubble before and after crossing the saddle point as the critical bubble, some authors~\cite{Punnathanam03,Uline07,Uline08} studied a constrained density functional model that were defined not with a simple smooth density profile but rather by a fixed particle number contained within a given small volume.  It is not clear if the appearance of the negative eigenvalues for the stability matrix of this constrained system means the unstable growing mode of the growing bubble after crossing the saddle point.

Finally, our stability analysis using the Schr\"odinger equation can be applicable to the special case of the density functional model for the Yukawa fluid as the density functional for the Yukawa fluid can be transformed exactly into the square-gradient functional~\cite{Sullivan79,Iwamatsu95}.  The stability analysis of critical nucleus of this Yukawa model fluid will be presented elsewhere.

\begin{acknowledgments}
This work is supported in part by the Grant-in-Aid for Scientific Research (C)22540422 from Japan Society for the Promotion of Science (JSPS). This work was conducted during MI's sabbatical leave to Tokyo Metropolitan University (TMU) from Tokyo City University (TCU).  MI is grateful to Department of Physics, TMU, and Professor Y. Okabe for their hospitality, and TCU for the support to his sabbatical leave.  
\end{acknowledgments}

\end{document}